\documentclass[11pt]{article}
\usepackage[a4paper]{geometry}
\usepackage[english]{babel}

\usepackage{amsmath, accents}
\usepackage{amsfonts}
\usepackage{amstext}
\usepackage{amssymb}
\usepackage{amsthm}
\usepackage{amscd}
\usepackage{mathrsfs}

\usepackage{fancyhdr}
\usepackage[pagebackref,draft=false]{hyperref}
\hypersetup{colorlinks,
linkcolor=myrefcolor,
citecolor=mycitecolor,
urlcolor=myurlcolor}

\usepackage[capitalize]{cleveref}
\usepackage{cite}
\usepackage{caption}
\usepackage{etaremune}

\usepackage{xcolor}
\definecolor{myurlcolor}{rgb}{0,0,0.4}
\definecolor{mycitecolor}{rgb}{0,0.5,0}
\definecolor{myrefcolor}{rgb}{0.5,0,0}
\usepackage{graphicx}
\usepackage{tikz}
\usepackage{tikz-cd}
 \usetikzlibrary{decorations.pathmorphing}

\usepackage{etoolbox}
\usepackage{makeidx}
\usepackage{sectsty}
\usepackage{enumitem} 
\usepackage[]{latexsym}
\usepackage{braket}
\usepackage{caption}
\usepackage[utf8]{inputenx}
\usepackage[T1]{fontenc}
\usepackage{lmodern}
\usepackage{textcomp}
\usepackage{microtype}
\usepackage{totcount}
\usepackage{blindtext}

\newtheorem*{proof*}{Proof}

\newcommand{\be}{\begin{equation}}
\newcommand{\ee}{\end{equation}}
\newcommand{\bea}{\begin{eqnarray}}
\newcommand{\eea}{\end{eqnarray}}
\newcommand{\vsp}{\vspace{0.4cm}}

\newcommand{\hh}{\mathcal{H}}
\newcommand{\bh}{\mathcal{B}(\mathcal{H})}
\newcommand{\tbh}{\mathbf{T}\mathcal{B}(\mathcal{H})}

\newcommand{\Uh}{\mathcal{U}(\mathcal{H})}
\newcommand{\uh}{\mathfrak{u}(\mathcal{H})}

\newcommand{\lag}{\mathfrak{L}}

\title{Lagrangian description of Heisenberg and Landau-von Neumann equations of motion}
\date{}

\author{F. M. Ciaglia$^{1,7}$ \href{https://orcid.org/0000-0002-8987-1181}{\includegraphics[scale=0.7]{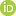}}, F. Di Cosmo$^{2,3,8}$ \href{https://orcid.org/0000-0003-0256-5913}{\includegraphics[scale=0.7]{ORCID.png}}, A. Ibort$^{2,3,9}$ \href{https://orcid.org/0000-0002-0580-5858}{\includegraphics[scale=0.7]{ORCID.png}}, \\ G. Marmo$^{4,5,10}$ \href{https://orcid.org/0000-0003-2662-2193}{\includegraphics[scale=0.7]{ORCID.png}}, L. Schiavone$^{3,4,6,11}$  \href{https://orcid.org/0000-0002-1817-5752}{\includegraphics[scale=0.7]{ORCID.png}}, A. Zampini$^{4,6,12}$ \href{https://orcid.org/0000-0003-0980-6003}{\includegraphics[scale=0.7]{ORCID.png}} \\
\footnotesize{$^{1}$\textit{ Max Planck Institute for Mathematics in the Sciences, Leipzig, Germany}} \\
\footnotesize{$^{2}$\textit{ ICMAT, Instituto de Ciencias Matem\'{a}ticas (CSIC-UAM-UC3M-UCM)}} \\
\footnotesize{$^{3}$\textit{Depto. de Matem\'aticas, Univ. Carlos III de Madrid, Legan\'es, Madrid, Spain}} \\
\footnotesize{$^{4}$\textit{ INFN-Sezione di Napoli, Naples, Italy}} \\
\footnotesize{$^{5}$\textit{ Dipartimento di Fisica ``E. Pancini'', Universit\`a di Napoli Federico II,  Naples, Italy}} \\
\footnotesize{$^{6}$\textit{ Dipartimento di Matematica e Applicazioni "Renato Caccioppoli", Università di Napoli Federico II, Napoli, Italy}} \\
\footnotesize{$^{7}$\textit{ e-mail: \texttt{florio.m.ciaglia[at]gmail.com} and \texttt{ciaglia[at]mis.mpg.de}}} \\
\footnotesize{$^{8}$\textit{ e-mail: \texttt{fcosmo[at]math.uc3m.es}}} \\
\footnotesize{$^{9}$\textit{ e-mail: \texttt{albertoi[at]math.uc3m.es}}} \\
\footnotesize{$^{10}$\textit{ e-mail: \texttt{marmo[at]na.infn.it}}} \\ 
\footnotesize{$^{11}$\textit{ e-mail: \texttt{luca.schiavone[at]unina.it}}} \\
\footnotesize{$^{12}$\textit{ e-mail: \texttt{alessandro.zampini[at]unina.it}}} 
}

\begin{document}

\maketitle

\begin{abstract}
An explicit Lagrangian description is given for the Heisenberg equation on the algebra of operators of a quantum system, and for the Landau-von Neumann equation on the manifold of quantum states which are isospectral with respect to a fixed reference quantum state.

\end{abstract}

\thispagestyle{fancy}

\section{Introduction}

Arguably, a Lagrangian description of quantum motion is a foundational requirement of any quantum mechanical description of physical systems.
Indeed, the usual canonical Hamiltonian formulation, for instance, turns out to be unduly complicated to deal with relativistic quantum evolution.
Following a suggestion by Dirac, \cite{dirac-the_lagrangian_in_quantum_mechanics} both Feynman, \cite{feynman_hibbs-quantum_mechanics_and_path_integrals} and Schwinger, \cite{Schwinger-2000} proposed their own Lagrangian formulations with far-reaching consequences in the description of quantum electrodynamics in particular, and quantum mechanics in a broad sense.

While a Lagrangian description on the Hilbert space associated with any quantum system has been considered already at the dawn of quantum Mechanics by  Frenkel in Ref. \cite{Frenkel-1934}, a Lagrangian description for Heisenberg algebraic setting  of the equations of motion has not been dealt with.  Indeed, the approach by Feynman, \cite{feynman_hibbs-quantum_mechanics_and_path_integrals} could be considered an ``integral approach'' by means of transition probabilities along histories. 
The Schwinger's approach, \cite{Schwinger-2000} again started from transition probabilities between prescribed initial and final states in describing the history as in the standard formalism of the S-matrix, to arrive at a ``quantum action principle'' in terms of a Lagrangian operator. The role of this Lagrangian operator formalism has been partially unveiled in the recent groupoidal description of Schwinger's formalism, \cite{Ci1,Ci2,Ci3,Ci4,Ci5,Ci6} where it was shown to describe a relevant family of quantum states.

In this letter, however, we would like to follow a different approach and, inspired by the geometrical formulation of quantum mechanics  (see, for instance Ref.\cite{A-S-1999,C-CG-M-2007,C-I-M-M-2015}), we will use geometrical methods, \cite{Morandi_Ferrario_LoVecchio_Marmo_Rubano_The_inverse_problem_in_the_Calculus_of_Variations_and_the_geometry_of_the_tangent_bundle} to provide a Lagrangian description  of quantum evolution. Specifically, we will provide a Lagrangian description for the Heisenberg and for the Landau-von Neumann equations.

In the Schr\"odinger picture of quantum mechanics, we have a complex Hilbert space $\mathcal{H}$ associated with a quantum system, and the dynamical evolution of a closed system is governed by the Schr\"odinger equation
\be \label{Eq: schrodinger}
i \frac{\mathrm{d}}{\mathrm{d}t} |\psi \rangle \,=\, \mathbf{H} |\psi \rangle \,,
\ee 
where $\mathbf{H}$ is a self-adjoint operator on $\mathcal{H}$. 
A Lagrangian formulation of this equation can be given by means of the Lagrangian function (see Ref.\cite{Frenkel-1934,Kramer_Saraceno_Geometry_of_the_time_dependent_variational_principle_in_Quantum_Mechanics,Tronci_Luz-Quantum_variational_principles})
\be \label{Eq: lagrangian schrodinger}
\lag \,=\, \frac{i}{2} \left(\, \langle \psi | \dot{\psi} \rangle - \langle \dot{\psi} | \psi \rangle \, \right) - \frac{1}{2} \langle \psi | \,\mathbf{H}\, | \psi \rangle \, .
\ee
This Lagrangian function may now be pulled-back to the tangent bundle of any submanifold of trial states.
This procedure is particularly useful in quantum information, where usually one considers a parametrized set of probability distributions, for instance Gaussian ones, and one is interested in the existence of additional geometrical structures.
The Lagrangian function is then used to define a distinguishability/divergence function which in turn may be used to define a metric tensor and a dualistic connection \cite{Ciaglia_DiCosmo_Hamilton_Jacobi_Potential_Functions,C-DC-M-2017, a_pedagogical}.

Starting from the equations of motion and seeking for a Lagrangian description is usually known as the ``inverse problem of the calculus of variations''.
The problem was first considered in a clear-cut mathematical formulation by Helmholtz in Ref.\cite{Helmholtz_inverse_problem}, while, in a related form, was posed by Wigner  in Ref.\cite{Wigner-1950} in the framework of quantum mechanics, and considered also by Feynman, as reported by Dyson after Feynman's death \cite{Ca95}.   An extension of the inverse problem to deal with a Lagrangian formulation for a  class of coupled dynamical systems, where one system obeys a second order, ordinary differential equation, while the second one obeys a first order, ordinary differential equation can be found in Ref.\cite{Ibort_Solano_On_the_inverse_problem_of_the_calculus_of_variations_for_a_class_of_coupled_dynamical_systems}.
Here we shall not deal with this problem in any generality, and we limit ourselves to provide explict Lagrangian descriptions for  the Heisenberg equation and the Landau-von Neumann equation.

In a mechanical-like context, the Lagrangian description of a dynamical evolution takes place on the tangent bundle of a given configuration manifold, and mainly deals with second order, ordinary differential equations (implicit or explicit).
However, in what follows we will be considering first order ordinary differential equations and we will present a preliminary analysis of this problem focused on the evolution equations of quantum mechanics.
It is worth to mention that, in the context of field theories, the problem has been first posed  in  Ref.\cite{Rossi_Reduction_order_theorem} and then solved in Ref.\cite{Saunders_reduction_order}, but the methods applied there make extensive use of a field theoretical framework, and necessarily require the equations considered to be genuine partial differential equations.
A Lagrangian procedure has been also proposed in Ref.\cite{Tronci_Luz-Quantum_variational_principles}, where the authors adopt the Euler-Poincar\`{e} reduction approach to write the equations of motion on the dual of the Lie algebra. 
However, the approach we follow here is conceptually different since  we will make direct use of traditional Lagrangian methods, together with the Dirac-Bergmann constraints analysis.

\vsp

\section{Heisenberg picture}

In the Heisenberg picture of quantum mechanics the relevant carrier space is the space of bounded linear operators $\bh$ on $\hh$.
The dynamical evolution of a closed quantum system is then described by a one-parameter group of algebra automorphisms given by
$A_{t}\,=\,\mathbf{U}_{t}^{\dagger}\,A\,\mathbf{U}_{t}$,
where $A\in\bh$ and $\mathbf{U}_{t}$ is the one-parameter group of unitary operators
$\mathbf{U}_{t}\,=\,\mathrm{e}^{- i t \mathbf{H}}$,
with $\mathbf{H}$ a self-adjoint operator on $\hh$, namely, the Hamiltonian of the system.
The infinitesimal version of the dynamical evolution is given by the Heisenberg equation 
\be\label{eqn: Heisenberg equation}
i\,\dot{A}\,\equiv\,i \frac{\mathrm{d}}{\mathrm{d}t} A \,=\, [A,\, \mathbf{H}].
\ee

We want to write down a Lagrangian function $\lag$ on the tangent bundle $\mathbf{T}\bh$ of $\bh$, whose Euler-Lagrange equations are equivalent to Heisenberg equation (\ref{eqn: Heisenberg equation}).
We consider a real-valued Lagrangian function $\lag$ which is linear in the velocities given by:
\be
\lag  \,=\, \frac{i}{2} \mathrm{Tr}\left(\, A^\dag \dot{A} - \dot{A}^\dag A \,\right)  -  \mathrm{Tr}\left( \, A \mathbf{H} A^\dag - A^\dag \mathbf{H} A \, \right),
\label{Lagrangian_function_Heisenberg}
\ee
where $\mathrm{Tr}$ is the canonical trace on $\bh$.
Here, we made implicit use of the trivialization $\tbh\cong\bh\times\bh$ for the tangent bundle of $\bh$ (which follows from the fact that $\bh$ is a  vector space). 
Then, an element in $\tbh$ is written as the couple $(A,\dot{A})$, where $\dot{A}$ indicates a tangent vector at $A$.

Let us remark now that the previous Lagrangian is defined on a subspace of the tangent bundle of an algebra of bounded operators on a Hilbert space $\mathcal{H}$. 
Generalizing this point of view, it would be possible to extend such a Lagrangian to generic von Neumann algebras, with the Trace operator on $\mathcal{B}(\mathcal{H})$ replaced by a suitable tracial state and, as in the previous instance, the Lagrangian defined only on a subspace of operators for which the expectation value makes sense. 


In order to write the Euler-Lagrange equations associated with $\lag$,  we build the so called Poincar\'e-Cartan one-form $\theta_{\lag}$ which is given by:
\be
\theta_{\lag}\, =\,\frac{i}{2} \,\mathrm{Tr}\left(A^{\dagger}\mathrm{d}A - A\mathrm{d}A^{\dagger}\right)\,,
\ee
where $\mathrm{d}A$ is an operator-valued one-form. Note that $\mathrm{d}A$ and $\mathrm{d}A^{\dagger}$ are functionally independent since we consider  $\bh$ as a real vector space and the Trace operator acts naturally on $\mathrm{d}A$. 
The Cartan two-form $\omega_{\lag} = -\mathrm{d} \theta_{\lag}$ becomes:
\begin{equation}
\omega_{\lag} = i\, \mathrm{Tr}(\mathrm{d}A \wedge \mathrm{d}A^{\dagger})\,,
\end{equation}
and it identically vanishes on the space of bounded Hermitian operators in $\bh$ as well as the one-form $\theta_{\lag}$.
Nonetheless, the equations of motion resulting from $\lag$ are coherent with the quantum evolution, as we will see in a moment.

According to \cite[Eq. 1.8]{Morandi_Ferrario_LoVecchio_Marmo_Rubano_The_inverse_problem_in_the_Calculus_of_Variations_and_the_geometry_of_the_tangent_bundle}, the Euler-Lagrange equations defined by the Lagrangian $\lag$ can be written as:
\be
\frac{\mathrm{d}}{\mathrm{d}t}\,\theta_{\lag}\,=\,\mathrm{d}\lag\,,
\ee
where the time derivative  stands for the Lie derivative with respect to the vector field describing the dynamics. 
Bearing in mind that the time derivative commutes with the exterior differential, we get:
\be
\frac{\mathrm{d}}{\mathrm{d}t}\,\theta_{\lag}\,=\,\frac{i}{2} \,\mathrm{Tr}\left(\dot{A}^{\dagger}\mathrm{d}A + A^{\dagger}\mathrm{d}\dot{A} - \dot{A}\mathrm{d}A^{\dagger} - A\mathrm{d}\dot{A}^{\dagger}\right),
\ee
and
\be
\mathrm{d}\lag\,=\,\frac{i}{2} \mathrm{Tr}\left(\, \dot{A}\mathrm{d}A^\dag +A^{\dagger}\mathrm{d}\dot{A} - \dot{A}^\dag \mathrm{d}A - A \mathrm{d}\dot{A}^{\dagger} \,\right)  -  \mathrm{Tr}\left( \, \left[ \mathbf{H},\,A^{\dagger}\right]\,\mathrm{d}A + \left[ A\,,\,\mathbf{H}\right]\,\mathrm{d}A^{\dagger} \right),
\ee
and thus the Euler-Lagrange equations become:
\be\label{eqn: EL H}
\mathrm{Tr}\left(\left(i\dot{A}^{\dagger} +\left[\mathbf{H},\,A^{\dagger}\right]\right)\,\mathrm{d}A + \left(\left[A\,,\,\mathbf{H}\right] - i\dot{A}\right)\,\mathrm{d}A^{\dagger}\right)\,=\,0\,.
\ee
Since $\mathrm{d}A$ and $\mathrm{d}A^{\dagger}$ are independent,  for equation \eqref{eqn: EL H} to hold we must impose that their multiplying factors vanish.
Furthermore, if $\mathbf{H}$ is Hermitian, note that the expression multiplying $\mathrm{d}A $ is precisely the adjoint of the coefficient multiplying $\mathrm{d}A ^{\dagger}$, and thus we obtain that equation \eqref{eqn: EL H} is equivalent to Heisenberg equation \eqref{eqn: Heisenberg equation}. 
In other words, we have obtained a Lagrangian description for the Heisenberg equation passing from the space of Hermitian operator to its complexification, which is the ``unfolding space'' $\bh$ (in the spirit of the unfolding procedure as given, for instance, in Ref.\cite{C-I-M-M-2015}). 
We will see another instance of this phenomenon in the following section.

We end this section providing an example inspired by the so called ``parametric Information Geometry'', where the focus is on parameterised families of states forming a manifold.
Generalizing this idea, we will consider a parameterised family of elements of the algebra $\bh$, and we will induce a dynamical evolution on this set by restricting the Lagrangian we have introduced for the Heisenberg equation. 
For the sake of simplicity, we will consider a subset of observables obtained as the orbit of a group action on the algebra $\bh$. 
Being more specific, we will consider a two-level quantum system under the left action of the group $SB(2,\mathbb{C})$. 
We notice that this group is interesting because it emerges as a subgroup of $SL(2,\mathbb{C})$, the complexification of $SU(2)$ in the description of a qubit. 

Let $\bh = \mathbf{M}_2(\mathbb{C})$ be the algebra of two-by-two matrices with complex entries, and let $A_0\in \bh$ be a chosen reference element of the algebra. 
The group $SB(2,\mathbb{C})$ is the group of two-by-two upper triangular matrices with unit determinant, that can be parameterized as follows:
\be
SB(2,\mathbb{C}) \ni g = \left(  
\begin{array}{cc}
 r & x+iy\\
 0 & r^{-1}
\end{array}
\right)  \, , \qquad r\in \mathbb{R}_+, \quad x,y \in \mathbb{R} \, .
\ee
This group may act on $\bh$ from the left by matrix multiplication and defines the orbit:
\be
\mathcal{S}= \left\lbrace A(g) = g A_0\, , \, g \in SB(2,\mathbb{C}) \right\rbrace \, .
\ee

The Lagrangian formulation introduced for the Heisenberg equation can be used to induce a dynamics on the set $\mathcal{S}$ in analogy with what is done in Ref.\cite{Non_linear_dynamics}. 
Indeed, the pull-back of the Lagrangian function in Eq. $\eqref{Lagrangian_function_Heisenberg}$ to $\mathcal{S}$ defines  a Lagrangian function on $\mathbf{T} \mathcal{S}$ given by:
\be
\lag_S = \mathrm{Tr}\left( g\rho_0g^{\dagger} \mathrm{Im}(\dot{g}g^{-1}) \right) - \mathrm{Tr}\left( \mathbf{H} \left[ g A_0 , (g A_0)^{\dagger} \right] \right)\,,
\label{Lagrangian_fuinction_S}
\ee
where $\rho_0 = A_0A_0^\dagger$, and $\mathrm{Im}(\dot{g}g^{-1}) = \frac{i}{2}\left( \left( \dot{g}g^{-1} \right)^{\dagger} - \dot{g}g^{-1} \right)$. 

These equations can be analyzed also from the point of view of first-order Lagrangian theories on Lie groups, see, for instance, Ref.\cite{Delgado_T_llez_2015} where such dynamics was thoroughly described when the first order kinetic term of the Lagrangian is given by an invariant 1-form on the group.   However, as it can easily checked, the kinetic term of the Lagrangian function $\lag_S$ is not invariant.
In spite of this it is easy to check that the Cartan two-form $\omega_{\lag} = - \mathrm{d}\theta_{\lag}$ has a kernel containing the vector fields which are tangent to the fibers of the tangent bundle of the Lie group, and the generators of the isotropy algebra of $g \rho_0 g^\dagger$.   The existence of such additional degeneracy of the Cartan 2-form will have the consequence that the Lagrangian equations will give rise to a constraint.  We will proceed by a direct computation of the equaations of motion  associated to  $\lag_S$:
\be
\begin{split}
\frac{\mathrm{d}}{\mathrm{d}t} \theta_{\lag_S} - d\lag_S = \mathrm{Tr}\left(gA_0\left(iA_0^{\dagger}\dot{g}^{\dagger} +\left[\mathbf{H},\,(gA_0)^{\dagger}\right]\right)\,(\mathrm{d}g)g^{-1} +\right.\\
+ \left. \left(\left[gA_0\,,\,\mathbf{H}\right] - i\dot{g}A_0\right)\left( gA_0 \right)^{\dagger}\,\left( (\mathrm{d}g)g^{-1} \right)^{\dagger}\right)\,=\,0\,. 
\end{split}
\ee
We can rewrite them as 
\begin{eqnarray}
& i \mathrm{Re} \left( \left( \dot{g}g^{-1} \right) g\rho_0 g^{\dagger} \right) - \frac{1}{2} \left[ g\rho_0 g^{\dagger} , \mathbf{H} \right] = 0 \label{ELEGG_1}\\
& \mathrm{Im} \left( \left( \dot{g}g^{-1} \right) g\rho_0 g^{\dagger} \right) - \frac{1}{2} \left\lbrace \mathbf{H} , g \rho_0 g^{\dagger} \right\rbrace + gA_0\mathbf{H}A_0^{\dagger}g^{\dagger} = 0 \label{ELEGG_2}\,,
\end{eqnarray}
and the symbol $\left\lbrace \cdot, \cdot \right\rbrace$ denotes the anticommutator of matrices.
We notice that these equations, in general, are not linear as it will be shown in detail later on.
In order to simplify the expressions in terms of the chosen coordinates for the group, let us introduce some notations:
$$
\rho_0 = \left( 
\begin{array}{cc}
c & a+ib \\
a-ib & d
\end{array}
\right) \, , \,\,
\mathbf{H} = \left( 
\begin{array}{cc}
\gamma & \alpha+i\beta \\
\alpha-i\beta & \delta
\end{array}
\right) \, , \,\,
 A_0\mathbf{H}A_0^{\dagger} = \left( 
\begin{array}{cc}
h_3 & h_1+ih_2 \\
h_1-ih_2 & h_4
\end{array}
\right)\,.
$$
Then, Euler-Lagrange equations can be written as the system of first-order ordinary differential equations in terms of the parameter description of $SB(2, \mathbb{C})$ given by
\begin{eqnarray}\label{eqn: EOM SB(2,C)}
& \dot{y}a - \dot{x}b = (\gamma c - h_{11})r + (\gamma a - h_1)x + (\gamma b - h_2)y - \frac{\delta d}{r^3} - d \frac{(\alpha x + \beta y)}{r^2}\label{ELE_S-1}\\
& \dot{r}b + \dot{y}d = (\gamma a - h_1)r + (\gamma d - h_4)x + \frac{d\alpha}{r}\label{ELE_S-2} \\
& -(\dot{r}a + \dot{x}d) = (\gamma b - h_2) r + (\gamma d - h_4)y + \frac{d\beta}{r} \label{ELE_S-3}\,.
\end{eqnarray}
Notice that these equations can be displayed as matrix implicit differential equations:
\be
\mathfrak{A} \dot{X} = Y,
\ee
where  
\begin{equation}
\dot{X}= \left( \begin{array}{r} \dot{x}\\ \dot{y}\\ \dot{r} 
\end{array} \right)\,,\qquad
\mathfrak{A} = \left( 
\begin{array}{rrr}
-b & a & 0 \\
0 & d & b \\
-d & 0 & -a
\end{array}
\right)\,,
\end{equation}
and
\be
Y= \left( \begin{array}{c} (\gamma c - h_3)r + (\gamma a - h_1)x + (\gamma b - h_2)y - \delta d/ r^3 - d (\alpha x + \beta y) / r^2 \\ (\gamma a - h_1)r + (\gamma d - h_4)x + d\alpha/ r \\ (\gamma b - h_2) r + (\gamma d - h_4)y + d\beta / r
\end{array} \right)\,.
\ee
The matrix $\mathfrak{A}$ has rank two and the vector $K= \left( a, b , -d \right)$
spans its kernel.
From the matricial form of the system of equations, it straighforwardly follows that there is a constraint on the admissible configurations given by:
\begin{equation*}
\begin{split}
d \left[ (\gamma c - h_{11})r + (\gamma a - h_1)x + (\gamma b - h_2)y - \frac{\delta d}{r^3} - d \frac{(\alpha x + \beta y)}{r^2}\right] = \\
= \left[ a(\gamma a - h_1) + b (\gamma b -h_2) \right]r +
+ (a \alpha + b \beta)\frac{d}{r} + (\gamma d - h_4)(xa + y b)\,.
\end{split}
\end{equation*}
Therefore, one has to treat this dynamical system according to the Dirac-Bergmann procedure. 
Let us see what is the final result in a simpler case, where the reference matrix $A_0$ is real and symmetric and $\mathbf{H}$ has only real entries.
In this particular instance, we have that $\beta = \alpha$, $b = 0$ and $h_2=0$, and the system of equations \eqref{eqn: EOM SB(2,C)} becomes:
\begin{eqnarray*}
& a \dot{y} = (\gamma c - h_3)r + (\gamma a - h_1 - \frac{d\alpha}{r^2})x - \frac{\delta d}{r^3} \\
& \dot{y}d = (\gamma a - h_1)r + (\gamma d - h_4)x + \frac{d\alpha}{r} \\
& (\dot{r}a + \dot{x}d) = - (\gamma d - h_4)y \,.
\end{eqnarray*}
From the first two equations one easily gets the  constraint:
\begin{equation*}
x \left[ (h_4a - d h_1) r^2 - d^2 \alpha \right] = \frac{1}{r}\left[ \left( a (\gamma a - h_1) - d(\gamma c - h_3) \right)r^4 + ad\alpha r^2 + (\delta d - h_4)d \right]\,.
\end{equation*}
If the coefficients are such that  $\left[ (h_4a - d h_1) r^2 - d^2 \alpha \right] \neq 0$, for all $r >0$,
the constraint can be explicitly solved and we have that the dynamics is restricted to the submanifold defined by the condition
\begin{equation*}
x = \frac{\left[ \left( a (\gamma a - h_1) - d(\gamma c - h_3) \right)r^4 + ad\alpha r^2 + (\delta d - h_4)d \right]}{r\left[ (h_4a - d h_1) r^2 - d^2 \alpha \right] } = \Phi(r)\,.
\end{equation*}
The dynamics is compatible with the previous constraint if 
\be
\dot{x} = \frac{\mathrm{d}\Phi}{\mathrm{d}r}\dot{r}\,.
\ee
Therefore, if we impose this additional constraint we obtain the  system of ordinary differential equations given by
$$
 \dot{y}d = (\gamma a - h_1)r + (\gamma d - h_4)\Phi(r) + \frac{d\alpha}{r} \, , \qquad   \dot{r}( a + \frac{\mathrm{d}\Phi}{\mathrm{d}r}d) = - (\gamma d - h_4)y\,.
$$
One can immediately notice that the final equations are non-linear, a phenomenon which typically occurs when we consider effective dynamics on subsets of states of interest parameterized by points of manifolds.
We shall not discuss possible physical applications of these equations here.

\section{Landau-von Neumann equation}
\label{sec.3}

At this point, having a Lagrangian description for the Heisenberg equation, we may want to provide a Lagrangian description also for the Landau-von Neumann equation
\be\label{eqn: lvN}
\dot{\rho}\,\equiv\,\frac{\mathrm{d}}{\mathrm{d}t}\,\rho \,=\,i\,\left[\rho\, ,\mathbf{H}\right]\,,
\ee
where $\rho$ is a quantum state (density operator on $\hh$).
This equation is  the infinitesimal version of the dynamical evolution given by
$\rho_{t}\,=\,\mathbf{U}_{t}\,\rho\,\mathbf{U}_{t}^{\dagger}$.

We remark, at this point, that Landau-von Neumann equation should not be understood as a mere ``dualization'' of equation \eqref{eqn: Heisenberg equation}. 
Indeed, while Heisenberg equation is defined for every Hermitian matrix, Landau-von Neumann equation is an equation on the space of states, which are positive and normalized density matrices. 
On the other hand, if we pull-back the Lagrangian function \eqref{Lagrangian_function_Heisenberg} to a manifold of states, seen as density operators on the Hilbert space $\mathcal{H}$ of the system, the associated Lagrangian one-form would identically vanish since it vanishes for all Hermitian matrices.

The dynamical evolution \eqref{eqn: lvN} takes place on the manifold $\mathcal{O}_{\sigma}$ of quantum states (density operators) which are   isospectral with respect to a fixed reference quantum state $\sigma$.
The manifold $\mathcal{O}_{\sigma}$ is a homogeneous space of the Lie group $\Uh$ of unitary operators on $\hh$, specifically, $\mathcal{O}_{\sigma}\cong\Uh / \Uh_{\sigma}$ where $\Uh_{\sigma}$ is the isotropy subgroup of $\sigma$ with respect to the action
\be\label{eqn: isospectral orbit}
(\mathbf{u},\sigma)\,\mapsto\,\mathbf{u}^{\dagger}\,\sigma\,\mathbf{u}\,.
\ee

In order to give a Lagrangian description for the Landau-von Neumann equation, we will exploit the fact that every orbit $\mathcal{O}_{\sigma}$ is a homogeneous space for the   unitary group $\Uh$.
Specifically, we first define a non-trivial immersion $\varphi$ of the unitary group $\Uh$ within $\bh$, and then, we  take the pullback $\lag_{u}$ of $\lag$ to $\mathbf{T}\Uh$ through the tangent map $T\varphi$, and consider the associated dynamical evolution.
It will be shown that the   Lagrangian two-form associated with $\lag_{u}$ will have a non-empty kernel, and that the the dynamics defined by it will actually be  defined on the orbit $\mathcal{O}_{\sigma}$ and coincide with the Landau-von Neumann equation.
This instance is particularly relevant when we recall again that the pullback to any submanifold of Hermitean operators of the Lagrangian two-form associated with \eqref{Lagrangian_function_Heisenberg} identically vanishes. 

In particular, we consider the map
\be
\varphi_{\sigma}(\mathbf{u})\,:=\,\sqrt{\sigma}\,\mathbf{u}
\label{immersion_L-vN}
\ee
where $\sigma$ is the reference quantum state defining the orbit $\mathcal{O}_{\sigma}$.
The map \eqref{immersion_L-vN} can be also interpreted as a way to associate to any positive operator $\sigma$ its ``operatorial complex square root''. 
Indeed, we can identify a module (the square-root of the positive operator) and a set of ``phases'' (the unitary operators). 
In this way, we obtain a new set of coordinates, a sort of polar coordinates, for the matrices belonging to $\mathcal{B}(\mathcal{H})$. 
This ``complexification'', obtained via the action of the unitary group, allows us to pull-back the Lagrangian function in equation \eqref{Lagrangian_function_Heisenberg} to the unitary group so that we obtain now the Lagrangian function, which now is of the form discussed in Ref. \cite{Delgado_T_llez_2015}:
\be
\lag_{u}\,=\,i\,\mathrm{Tr}\left( \sigma\, \dot{\mathbf{u}} \mathbf{u}^{\dagger}\right) - \,\mathrm{Tr}\left(\mathbf{u}^{\dagger}\sigma\mathbf{u}\,\mathbf{H} - \sigma\,\mathbf{H} \right) \,.
\ee


We will derive explicitly the equations of the Lagrangian function $\lag_{u}$ to show that it gives rise to the dynamics of the Landau-von Neumann equation on the manifold of quantum states states which are isospectral with respect to $\sigma$.  To achieve this we will need to recall a few facts from the differential geometry of the tangent bundle of Lie groups (see also, Ref.\cite{M-R-1988,Delgado_T_llez_2015}).

First of all, the elements $\Theta^{L}\equiv\mathbf{u}^{\dagger}\mathrm{d}\mathbf{u}$ and $\Theta^{R}\equiv \mathrm{d}\mathbf{u}\mathbf{u}^{\dagger}$ will denote the left and right-invariant Maurer-Cartan forms on $\mathcal{U}(\mathcal{H})$, respectively.
Then,  the Cartan one-form $\theta_{u}$ associated with $\lag_{u}$ takes the form:
\be
\theta_{u}\,=\,i\,\mathrm{Tr}\left(\sigma\,\Theta^{R}\right)\,=\,i\,\mathrm{Tr}\left(\sigma\,\mathrm{d}\mathbf{u}\,\mathbf{u}^{\dagger}\right).
\ee

Note that $\mathrm{d}\mathbf{u}$ and $\mathrm{d}\mathbf{u}^{\dagger}$ are not functionally independent because, since $\mathbf{u}\mathbf{u}^{\dagger}=\mathbb{I}$, we get:
\be
\mathbf{u}\mathrm{d}\mathbf{u}^{\dagger}\,=\,- (\mathrm{d}\mathbf{u})\,\mathbf{u}^{\dagger}.
\ee
Furthermore, a similar reasoning applies for $\dot{\mathbf{u}}$ and $\dot{\mathbf{u}}^{\dagger}$ leading to the equality $\mathbf{u}\,\dot{\mathbf{u}}^{\dagger}\,=\,- \dot{\mathbf{u}}\,\mathbf{u}^{\dagger}$.
Bearing these expressions in mind, we proceed to compute the Euler-Lagrange equations associated with $\lag_{u}$ as we did in the previous section for the Heisenberg equation and obtain:
\be
\frac{\mathrm{d}}{\mathrm{d}t}\theta_{u}\,=\,i\,\mathrm{Tr}\left(\mathbf{u}^{\dagger}\sigma\,\mathrm{d}\dot{\mathbf{u}} - \mathbf{u}^{\dagger}\dot{\mathbf{u}}\mathbf{u}^{\dagger}\sigma\mathbf{u} \,\Theta^{L}\right),
\ee
and
\be
\mathrm{d}\lag_{u}\,=\,i\,\mathrm{Tr}\left(\mathbf{u}^{\dagger}\sigma\,\mathrm{d}\dot{\mathbf{u}} - \mathbf{u}^{\dagger}\,\sigma\,\dot{\mathbf{u}}\, \,\Theta^{L}\right) - \mathrm{Tr}\left(\left[\mathbf{u}^{\dagger}\sigma\mathbf{u},\,\mathbf{H}\right]\,\Theta^{L}\right)\, ,
\ee
that, together, will provide for the Euler-Lagrange equations,  $\frac{\mathrm{d}}{\mathrm{d}t}\theta_{u}\,=\,\mathrm{d}\lag_{u}$, the expression:
\be
 i\,\mathrm{Tr}\left(  \left(\mathbf{u}^{\dagger}\,\sigma\,\dot{\mathbf{u}}\, - \mathbf{u}^{\dagger}\dot{\mathbf{u}}\mathbf{u}^{\dagger}\sigma\mathbf{u}  -  i\,\left[\mathbf{u}^{\dagger}\sigma\mathbf{u},\,\mathbf{H}\right]\right) \,\Theta^{L}\right) \,=\,0\,.
\ee
Now, if $\rho\,\equiv\,\mathbf{u}^{\dagger}\,\sigma\,\mathbf{u}$,
denotes a generic point in the manifold $\mathcal{O}_{\sigma}$, and we note that:
\be
\dot{\rho}\,\equiv\,\frac{\mathrm{d}}{\mathrm{d}t}\,\rho\,=\,\mathbf{u}^{\dagger}\,\sigma\,\dot{\mathbf{u}}\, - \mathbf{u}^{\dagger}\dot{\mathbf{u}}\mathbf{u}^{\dagger}\rho\mathbf{u},
\ee 
we can write the Euler-Lagrange equations as:
\begin{equation}\label{elun}
 i\,\mathrm{Tr}\left(  \left(\dot{\rho}  - i\,\left[ \rho ,\,\mathbf{H}\right]\right) \,\Theta^{L}\right) \,=\,0\,.
\end{equation}
The left-invariant Maurer-Cartan form $\Theta^{L}$ can also be written as  $\Theta^{L}\,=\,\tau_{j}\,\theta^{j}$, 
where $\{\tau_{j}\}_{j=1,...,n^{2}}$ is a basis of the Lie algebra of $\uh$ (which is orthonormal with respect to the Hilbert-Schmidt product), and $\theta^{j}$ is the left-invariant one-form on $\mathcal{U}(\mathcal{H})$ which is dual to the left-invariant vector field $X_{j}$ associated with the Lie algebra element $\tau_{j}$ (see Ref.\cite{M-R-1988,Delgado_T_llez_2015}).
Then, we write the Euler-Lagrange equations (\ref{elun}) as:
\be
 i\,\mathrm{Tr}\left(  \left(\dot{\rho} - i\,\left[ \rho ,\,\mathbf{H}\right]\right) \,\tau_{j}\right)\,\theta^{j} \,=\,0\,.
\ee
Since both the $\theta^{j}$'s and $\tau_j$'s are independent, the previous equation is equivalent to the Landau-von Neumann operator equation (\ref{eqn: lvN}) as claimed.

\vsp
Going back to the example of the group $SB(2,\mathbb{C})$, dual to $SU(2)$, discussed in the previous section, notice that if we consider $A_0=\sqrt{\sigma}$, with $\sigma$ a positive matrix, we obtain an evolution of the type $g(t)\sqrt{\sigma}$, with $g(t)$ solution of the equations \eqref{ELEGG_1}-\eqref{ELEGG_2}. 
Hence, upon normalization we obtain two dynamics on the space of states of a qubit:
\begin{equation}\label{eqn: two dynamics on qubit}
\rho_1(t) = \frac{g(t)\sigma g^{\dagger}(t)}{\mathrm{Tr}\left( g(t)\sigma g^{\dagger}(t) \right)} \,, \qquad
 \rho_2(t) = \frac{\sqrt{\sigma}g^{\dagger}(t)g(t)\sqrt{\sigma}}{\mathrm{Tr}\left( \sigma g^{\dagger}(t)g(t) \right)}\,.
\end{equation} 
These dynamical evolutions have a very different nature  with respect to the solutions of Landau-von Neumann equation \eqref{eqn: lvN}. 
In particular, we note that the dynamics $\rho_1(t)$ is associated with a nonlinear adjoint action of $SB(2,\mathbb{C})$ on the space of states of a qubit, while $\rho_2(t)$ is associated with an action of $SB(2,\mathbb{C})$ on the space of operators.

We want to conclude this section illustrating some features of the orbits $\mathcal{O}_{\sigma}$, given by the linear action \eqref{eqn: isospectral orbit}, and the non-linear $\tilde{\mathcal{O}}_{\sigma}=\left\lbrace g \sigma  g^{\dagger}/ \mathrm{Tr}(g \sigma g^{\dagger})  \left|\right. g\in SB(2,\mathbb{C}) \right\rbrace$, in the case of a qubit. 

It is well-known that the coadjoint orbits of $SU(2)$ on the algebra of positive matrices are diffeomorphic to bidimensional spheres. 
The vector field associated with Landau-von Neumann equation is tangent to this orbit and defines a Hamiltonian vector field on it (see for instance Ref.\cite{C-DC-I-L-M-2017}). 

In order to understand some features of the orbit $\tilde{\mathcal{O}}_{\sigma}$ we consider the generators of the one-parameter groups of transformation given by:
$$
\rho(t) = \frac{g_k(t)\sigma g_k^{\dagger}(t)}{\mathrm{Tr}(g_k(t)\sigma g_k^{\dagger}(t))}\,, \quad k = 1,2,3\,,
$$
where $g_k(t) = \exp t\tau_k$, and, $\tau_1 =  \left( 
\begin{array}{cc}
0 & 1 \\
0 & 0
\end{array}
 \right)$, $\tau_2 =\left( 
\begin{array}{cc}
0 & i \\
0 & 0
\end{array}
 \right)$, $\tau_3 = \left( 
\begin{array}{cc}
1 & 0 \\
0 & -1
\end{array}
 \right)$, is a basis of the Lie algebra of $SB(2,\mathbb{C})$.
The matrix $\sigma = \frac{1}{2}\left( I + \mathbf{x}\cdot \boldsymbol{\sigma} \right)$, $\mathbf{x} = (x_{1},x_{2},x_{3})$ and $r = || \mathbf{x} || \leq 1$, 
represents the most general state of a qubit. 
The infinitesimal generators of these one-parameter groups of transformations are the vector fields:
\begin{equation*}
\begin{split}
Y_1 = x_1\frac{\partial }{\partial x_3} -x_3\frac{\partial}{\partial x_1} + \frac{\partial}{\partial x_1} - x_1 \Delta \,\, , \, \, 
Y_2 = -x_2\frac{\partial }{\partial x_3} +x_3\frac{\partial}{\partial x_2} - \frac{\partial}{\partial x_2} + x_2 \Delta \,,\,\,\,
Y_3 = \frac{\partial }{\partial x_3} - x_3\Delta\,,
\end{split}
\end{equation*}
where $\Delta = x_1\frac{\partial}{\partial x_1} + x_2 \frac{\partial}{\partial x_2}+ x_3 \frac{\partial}{\partial x_3}$.
The wedge product of these vector fields gives the following multivector:
\begin{equation*}
Y_1\wedge Y_2 \wedge Y_3 = -(1-x_3)^2(1-r^2)\frac{\partial}{\partial x_1} \wedge \frac{\partial}{\partial x_2} \wedge \frac{\partial}{\partial x_3}\,,
\end{equation*}
which is only vanishing on the sphere of pure states ($r =1$).
Therefore, we can distinguish three types of integral leaves of the distribution $\left\lbrace Y_1,Y_2,Y_3 \right\rbrace$: a fixed point $P\equiv (0,0,1)$, its complement in the sphere of pure states, and the bulk of faithful (invertible) states. 
This last orbit is three dimensional and the vector fields $Y_1,Y_2,Y_3$ are never tangent to the coadjoint orbit of the unitary group: they move a point from one orbit to another one, which means they do not preserve the spectrum of a matrix, but they do preserve its determinant.
However, at any point of the bulk, it will be possible to express two generators of the coadjoint action of the unitary group as a combination of $Y_1$ and $Y_2$ having functions as coefficients. 


From these observations, we can already get some conclusions. 
The solutions of Heisenberg equation are one-parameter groups of unitary transformation acting via the adjoint action, and, in the case of a qubit, these flows lie on a sphere. 
The orbits of the group $SB(2,\mathbb{C})$, on the other hand,  live either on the sphere of pure states, or in the bulk of faithful (invertible) states, and the spectra are not preserved under the flow of the vector fields $Y_1,Y_2,Y_3$. 
Therefore, the Euler-Lagrange equations obtained from the pullback to $SB(2,\mathbb{C})$  of the Lagrangian in \eqref{Lagrangian_function_Heisenberg}, and then projected to the space of states, cannot have as a solution a one-parameter group of transformations which is a subgroup of $SB(2,\mathbb{C})$. 
When moving to higher dimensions, the situation will be more complicated and the dependence on the starting state $\sigma$ more acute.

\section{Conclusions}

Motivated by previous works on the Lagrangian description of Schr\"{o}dinger equation, we provided a Lagrangian description of the Heisenberg and of the Landau-von Neumann equations.
In particular, we  have shown how the Lagrangian description of the Landau-von Neumann equation may be obtained from the Lagrangian description of the Heisenberg equation by means of the pullback procedure introduced in Ref.\cite{Non_linear_dynamics}. 
This is in line with the recent developments of classical and quantum information geometry in which the focus is on parametrized submanifold of classical and quantum states rather than on the whole space of states.
Indeed, the Lagrangian approach, being formulated in terms of covariant objects, behaves naturally with respect to immersions, unlike first order differential equations like the Schr\"{o}dinger and the Heisenberg equations as pointed out, for instance, in Ref. \cite{Ciaglia_DiCosmo_Marmo_Ibort_Dynamical_aspects_in_the_QuantizedDequantizer_formalism}.

Having considered here only closed quantum systems, a reasonable next step to take is   to look for a Lagrangian description of open quantum systems in terms of the geometrical formalism introduced for instance in  Ref.\cite{C-DC-I-L-M-2017, boundary}. 
However, in dealing with dissipative systems, the Lagrangian description to look for should be given according to the formalism developed in Ref.\cite{deritis_marmo_platania_scudellaro-inverse_problem_in_classical_mechanics_dissipative_systems}, or building on the contact formalism given in Ref.\cite{bravetti_cruz_tapias-contact_hamiltonian_mechanics,C-C-M-2018}.

At the end of section \ref{sec.3}, we have also considered the possibility of defining non unitary and non linear dynamics on the space of states of a qubit via the pullback to a set of matrices parameterized by the group $SB(2,\mathbb{C})$ of the Lagrangian \eqref{Lagrangian_function_Heisenberg}, and then projecting the solutions to the space of states. 
A deeper analysis of all these aspects, however, will be addressed in future works.


\section*{Acknowledgments}

F.D.C. and A.I. would like to thank partial support provided by the MINECO research project MTM2017-84098-P and QUITEMAD++, S2018/TCS-A4342. A.I. and G.M. acknowledge financial support from the Spanish Ministry of Economy and Competitiveness, through the Severo Ochoa Programme for Centres of Excellence in RD(SEV-2015/0554). G.M. would like to thank the support provided by the Santander/UC3M Excellence Chair Programme 2019/2020, and he is also a member of the Gruppo Nazionale di Fisica Matematica (INDAM), Italy.


\addcontentsline{toc}{section}{References}
\bibliographystyle{plain}

\end{document}